\documentclass[]{spie}  

\usepackage{amsmath,amsfonts,amssymb}
\usepackage{graphicx}
\usepackage[colorlinks=true, allcolors=blue]{hyperref}
\usepackage{multirow}
\usepackage{enumitem} 
\usepackage[]{algorithm2e}

\usepackage{xcolor}
\usepackage{subfigure}
\usepackage[normalem]{ulem}
\usepackage{mathtools}
\usepackage{amsmath}
\usepackage{ulem}
\usepackage{txfonts}
\usepackage{textcomp} 
\usepackage{listings}

\title{Real-time control and data standardization on various telescopes and benches}

\author[a]{Nour Skaf}
\author[a]{Rebecca Jensen-Clem}
\author[a]{Aaron Hunter}
\author[b,d,e,f]{Olivier Guyon}
\author[b]{Vincent Deo}
\author[a]{Phil Hinz}
\author[a]{Sylvain Cetre}
\author[a]{Vincent Chambouleyron}
\author[a]{J. Fowler}
\author[a]{Aditya Sengupta}
\author[a]{Maissa Salama}
\author[e]{Jared Males}
\author[e]{Eden McEwen}
\author[e]{Ewan S. Douglas}
\author[e]{Kyle Van Gorkom}
\author[g]{Emiel Por}
\author[b]{Miles Lucas}
\author[h]{Florian Ferreira}
\author[h]{Arnaud Sevin}
\author[a]{Rachel Bowens-Rubin}
\author[i]{Jesse Cranney}
\author[j]{Ben Calvin}

\affil[a]{Department of Astronomy \& Astrophysics, University of California, Santa Cruz, CA 95064, USA}
\affil[b]{National Astronomical Observatory of Japan, Subaru Telescope, 650 North A'oh\=ok\=u Place, Hilo, HI 96720, U.S.A.}
\affil[d]{Astrobiology Center of NINS, 2-21-1 Osawa, Mitaka, Tokyo 181-8588, Japan}
\affil[e]{Steward Observatory, University of Arizona, Tucson, AZ 85721, USA}
\affil[f]{Wyant College of Optical Sciences, University of Arizona, Tucson, AZ 85721, USA}
\affil[g]{Space Telescope Science Institute (STScI), 3700 San Martin Dr, Baltimore MD, 21218, USA}
\affil[h]{LESIA, Observatoire de Paris, Universit\'e PSL, Sorbonne Universit\'e, Universit\'e de Paris, CNRS, 5 place Jules Janssen, 92195 Meudon, France}
\affil[i]{Research School of Astronomy and Astrophysics, Mt Stromlo Observatory, Weston Creek, ACT 2611, Australia}
\affil[j]{University of California - Los Angeles, 430 Portola Plaza, Los Angeles, CA 90095, USA}

\authorinfo{Further author information: (Send correspondence to N.S)\\.: E-mail: noskaf@ucsc.edu}

\begin{document} 
\maketitle

\begin{abstract}


Real-time control (RTC) is pivotal for any Adaptive Optics (AO) system, including high-contrast imaging of exoplanets and circumstellar environments. It is the brain of the AO system, and what wavefront sensing and control (WFS\&C) techniques need to work with to achieve unprecedented image quality and contrast, ultimately advancing our understanding of exoplanetary systems in the context of high contrast imaging (HCI). Developing WFS\&C algorithms first happens in simulation or a lab before deployment on-sky. The transition to on-sky testing is often challenging due to the different RTCs used. Sharing common RTC standards across labs and telescope instruments would considerably simplify this process. A data architecture based on the interprocess communication method known as shared memory is ideally suited for this purpose. 
The CACAO package, an example of RTC based on shared memory,  was initially developed for the Subaru-SCExAO instrument and now deployed on several benches and instruments.
This proceeding discusses the challenges, requirements, implementation strategies, and performance evaluations associated with integrating a shared memory-based RTC.
The Santa Cruz Extreme AO Laboratory (SEAL) bench is a platform for WFS\&C development for large ground-based segmented telescopes. 
Currently, SEAL offers the user a non-real-time version of CACAO, a shared-memory based RTC package initially developed for the Subaru-SCExAO instrument, and now deployed on several benches and instruments. We show here the example of the SEAL RTC upgrade as a precursor to both RTC upgrade at the 3-m Shane telescopes at Lick Observatory (Shane-AO) and a future  development platform for the Keck II AO. This paper is aimed at specialists in AO, astronomers, and WFS\&C scientists seeking a deeper introduction to the world of RTCs. 


\end{abstract}

\keywords{Adaptive Optics, Real-Time Control}

\section{INTRODUCTION}
\label{sec:intro}  

With nearly 6,000 discovered exoplanets and numerous dedicated space missions and ground-based instruments, the exoplanet era is advancing towards more robust detection and characterization methods. Different detection techniques, such as transit, radial velocity, direct imaging, and astrometry, are detecting new planets, enabling extensive statistical searches and surveys. However, direct imaging remains a significant technical challenge due to the overwhelming brightness of exoplanet host stars and the Earth's atmospheric distortions impacting the incoming wavefront.

In this context, AO has emerged as a pivotal technology for exoplanet science, offering more and more sophisticated wavefront sensing and control (WFS\&C) techniques to mitigate the deleterious effects of Earth's turbulent atmosphere, as well as the non-common path aberrations (NCPA), which are aberrations induced by the optical path between the AO loop and the science camera. Every high contrast imaging (HCI) instrument is equipped with an AO system, and the next generation of space-based telescopes will also be equipped with AO elements. Every AO system includes at least three core components: a wavefront sensor (WFS), an RTC, and a deformable mirror (DM). The RTC is comparable to the brain of AO: it is the software and hardware that performs the computation of the correction needed for the DM to correct atmospheric aberrations in real-time. In many cases, the RTC has been seen as a "set and forget" system, computing the corrections in the background of the science observations. However, interacting with an RTC for WFS\&C purposes is essential yet frequently perceived as complex by astronomers, requiring significant time and effort to adapt WFS\&C techniques to different RTC systems across various benches and telescopes. Furthermore, because the RTC is intrinsically part of the AO ecosystem, it is part of the experimental framework that must be flexible rather than static. 

In response to these challenges, some members of the RTC community have moved towards establishing common standards across instruments, notably through the use of shared-memory architectures for data management and interaction with the hardware. 
Mainly, the use of shared memory has become a standard in dealing with the data flows, which represents several benefits, especially in terms of modularity and flexibility. The use of shared-memories for RTC is originally from the CACAO (Compute and Control for Adaptive Optics) software \cite{Guyoncacao2020}, initially designed for the Subaru Coronagraph Extreme Adaptive Optics (SCExAO) \cite{SCExAO_Lozi_2018} needs. It is now being used by several additional instruments and optical benches. 
The target audience of this paper is AO scientists, astronomers, and WFS\&C scientists seeking a deeper introduction to the RTC world. We will focus especially on the RTC installation and integration using a few examples, focusing on the integration of CACAO on the Santa Cruz Extreme AO Lab (SEAL) testbed \cite{jensenclem2021seal}. We will describe the power of the use of shared-memories as a standard for data management architecture, as part of a long-term global strategic move towards a more collaborative and efficient scientific environment, in Sections \ref{sec:shm} and \ref{sec:shmrational}. 
Nowadays, many RTCs are based on CACAO's shared-memory architecture, each being designed for different uses, yet all benefiting from the modularity and flexibility of the shared-memory architecture which we will develop in Sections \ref{sec:shm} and \ref{sec:shmrational}. Although this paper focuses more on CACAO, we also note that other shared-memory-based RTCs are currently in use, some of which are described in Section \ref{sec:diffRTC}. 
The focus on CACAO in this paper stems from the on-going effort to develop CACAO as an open-source RTC industry standard. It is being upgraded with both community-oriented and performance-oriented goals, in order to facilitate adoption at new facilities. In this paper we use the SEAL testbed to demonstrate its implementation. In the process, we present some best-practices recommendations for successful integration and collaboration in the use of RTC systems. 
Finally, we lay out the vision of the opportunities for an integrated RTC system that includes SEAL, the Shane 3-m telescope (ShaneAO) at Lick Observatory, and Keck II AO, for both technical development and science observations. 



\section{Overview of the RTC}

\subsection{The place of the RTC in an AO system}
The AO system is composed of three main components: the wavefront sensor (WFS), that measures the incoming wavefront; the deformable mirror (DM), that corrects the incoming wavefront; the real-time control (RTC), that links the WFS to the DM.

Light averrated by earth atmosphere travels through the telescope system to the WFS. Once measured by the WFS, the wavefront aberrations must be reconstructed, allowing us to estimate the wavefront phase. The information from the WFS is sent to the RTC that reconstructs the wavefront phase. The RTC proceeds in an open, pseudo-open, or more generally closed loop operating in the kHz regime to keep up with the rapidly changing turbulence of the atmosphere. 
The RTC calculates the commands that need to be sent to the DM, which is the hardware element that compensates for the distorted wavefront. 
In a nutshell, the WFS first samples the wavefront, the RTC calculates its phase, and the DM acts accordingly to correct the aberrations. 

The RTC is the interface between the wavefront received and its correction. It plays two significant roles: estimating the wavefront based on the WFS data and then applying a relevant command to send to the DM. There is usually a linear relationship between these two components, allowing us to rely on linear algebra, with matrix-vector multiplication (MVM) to run the control loop. This must be done in real-time as the atmosphere is continuously evolving.

\subsection{Real-time operating system}

Over the past half-century, advancements in computer science and processor development have led to increasingly sophisticated operating systems (OS) with complex task management capabilities designed to maximize the efficient utilization of processor resources. This complexity, however, has rendered modern computers non-deterministic in their task management processes, causing them to perform tasks with variable latency and no guarantee of the order of execution. Real-time control systems, however, require deterministic (that is to say, fixed) latency to properly control the closed loop behavior of a dyanamic system such as an AO system. Modern computers are often fast enough that the variation in latency is small compared to the loop time, but AO systems increasingly require shorter loop times and more degrees of freedom to correct for atmospheric disturbances. These requirements stress the performance of the processor to the point where the latency variation affects the controller performance. 

To illustrate this point, we explore the concept of determinism. On a single CPU, the computer can execute only one process at a time. There is a scheduler that is responsible for allocating the required time for each process and determining which task to perform next. The scheduler is designed to manage task order autonomously, balancing all processes with equal priority to manage resources efficiently.
Because the scheduler manages many processes simultaneously, its behavior is challenging to predict. The scheduler selects which task to execute based on various parameters, such as the task's readiness, but importantly not according to a predetermined task priority. There is minimal visibility into the scheduler's decision-making process and task execution order can change instantaneously, contributing to the computer's non-deterministic nature.

One of the primary focuses of RTC development is to change the behavior of the core of the operating system to achieve deterministic behavior. 
Simplistically speaking, the RTC imposes specific priorities on tasks, thereby directing the scheduler to perform these tasks in order of importance. Less important tasks are preempted by more important tasks. By imposing these priorities, RTC systems achieve more deterministic behavior, ensuring predictable and reliable task execution essential for real-time control. For additional information on real-time kernel and performance, we refer to the SPIE Proceeding of Deo et al. 2024.

\subsection{Real-time control essentials}
In this section, we lay out the basics of what an RTC has to do, including key definitions and linear algebra concepts.  

\subsubsection{Matrices definitions}
Linear algebra is at the base of AO control\footnote{Interestingly, the literal translation of the Arabic word "Al-gebra" means "reunion of broken parts", introduced by the Persian mathematician Muhammad ibn Musa al-Khwarizma around 820 AD in his book "The Science of Restoring and Balancing". An etymological origin that perfectly fits AO, as well as the human analogy. The three centers must be balanced to make a healthy human being.}. There are three main matrices taking place in the control loop: 

\textbf{Response matrix:} The response matrix (RM) is the linear mapping from the WFS measurements $s$ to the DM commands $c$: $s=Dc$ ; with $D$ being the response matrix. It is usually measured by simply poking the DM with a
combination of modes and recording the response on the WFS.\\

\textbf{Control matrix:} Once the relationship between the WFS and the DM is acquired with the RM, we aim to estimate the DM commands to compensate for the measured WFS signal. To this end, we must compute the control matrix, which is usually the pseudo-inverse of the response matrix: $C=D^\dagger$. However, some modes applied on the DM during the RM acquisition have a weak signal on the WFS. These modes must be removed to avoid any excess noise. This is usually performed with a singular value decomposition (SVD) of the RM, and the rejection of the weak modes, before calculating its pseudo-inverse. \\

\textbf{Reconstruction matrix:} A modal reconstruction matrix translates the WFS measurements to modes. Different approaches include going from WFS to DM with a control matrix and going from WFS to modes with the reconstruction matrix, and then from modes to DM. This second approach is the modal control framework.\\

\subsubsection{Control loop}\label{sec:controlloop}
\begin{figure}[ht!]
  \centering
    \includegraphics[width=0.8\textwidth]{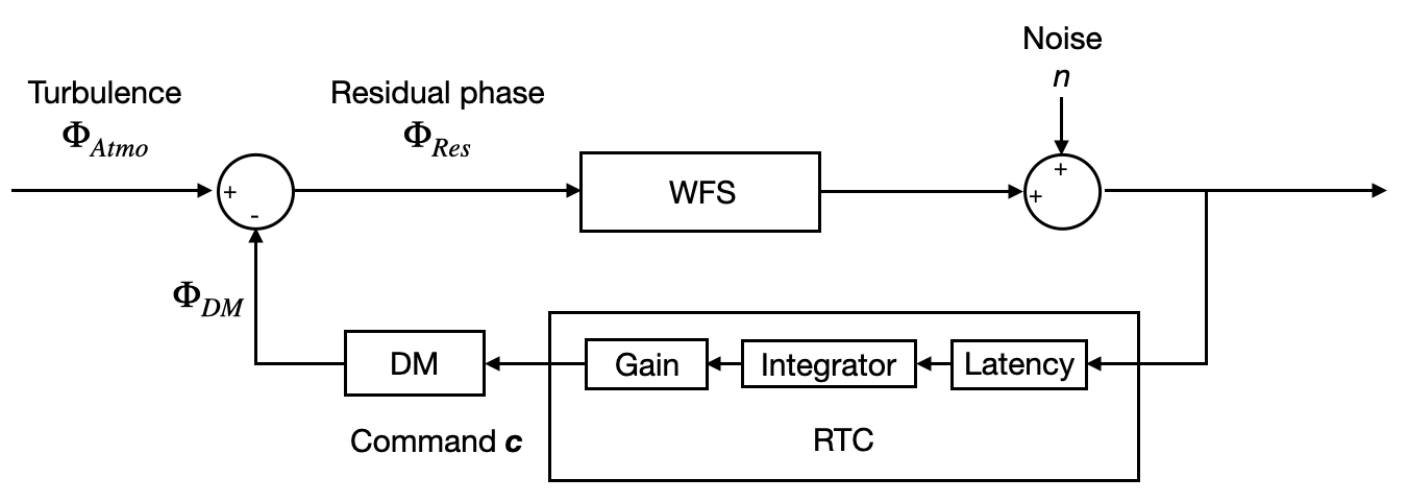}
      \caption{Simplistic block diagram of the control loop, in closed-loop, with an integrator filter.}
    \label{fig:RTC_loop}
\end{figure}

The control loop allows us to link WFS measurements to DM commands, and its temporal behavior is inherently part of the AO system's error budget. As shown in Figure \ref{fig:RTC_loop}, control laws ensure that the DM applies the relevant correction, taking into account the noise propagation and latency effects, while applying a gain to the correction
\cite{Demerle1994}. Typically, an integrator filter is introduced, so only a fraction of the correction, defined by the gain $g$, is applied to avoid instabilities linked to latency between WFS measurements and DM correction and to allow the reduction of photon noise by averaging the noisy WFS measurements. \\
We describe 
In the case of a closed loop, for a frame $k$, a perfectly corrected wavefront corresponds to: $ \Phi_{DM}[k] = \Phi_{Atm}[k]$;
and the WFS sees a perfectly flat wavefront. With the temporal evolution of the atmospheric turbulence, the WFS sees the frame $k+1$ with residual errors:
\begin{equation}
  \Phi_{Res}[k+1] = \Phi_{Atm}[k+1] - \Phi_{DM}[k] =  \Phi_{Atm}[k+1] - \Phi_{Atm}[k]
\end{equation}

There is hence a need to decrement the command $\Phi_{DM}$ of $\Phi_{Res}[k+1]$. We then need to use a command law with an integrator filter: for each frame calculated with the RTC, the command $c$ applied to the DM follows:
\begin{equation}
   c[k] = c[k-1] - g \times C \times s[k]
\end{equation}
with $s[k]$ being the WFS measurement at the frame $k$, $C$ being the control matrix, and $g$ being the gain of the integrator. This is a simplistic description of the control loop with its gain in the case of an integrator type of controller. The gain is usually lower than 0.5, to avoid any loop instability. In this case, that means that 50\% of the correction as measured by the WFS is applied. There are of course several other types of control laws besides the integrator, such as the Linear Quadratic Gaussian control \cite{Kulcsar2006}, the predictive control \cite{GuyonMales2017}\cite{VanKooten2022} \cite{Fowler2022}, and reinforcement learning \cite{Nousiainen2022RL} \cite{RLControlRecht}. 

The AO control loop includes several steps that need to be processed by the RTC. Representative steps of the control loop include: 
\begin{itemize}
    \item Wait for WFS image $s$;
    \item Calibrate: normalise WFS frame to the total intensity;
    \item Subtract the WFS reference;
    \item Compute and extract the WFS measurements;
    \item Multiply the resulting by the control matrix $C$;
    \item Update DM state, including loop gain $g$;
    \item Apply correction to DM;
    \item Record DM state for next loop iteration;
    \item Repeat until the loop opens.
\end{itemize}

Several remaining sources of error have a considerable impact on the final AO system performance at the science camera. They are characterized by the error variance they leave on the final phase, and the total variance of the residual phase corresponds to the sum of the individual variances, defining the error budget. 
There are errors due to the dimensioning of the system, mainly fitting and aliasing errors. Then there are errors due to the system itself, and the main error stemming from the RTC is the \textbf{temporal error}, linked with the lag between the measurement of the phase and its correction on the DM. Several studies expand on the AO error budget, such as the pure lag and the bandwidth error, and how these relate with the expose time and readouttime of the WFS detector, computation time on the RTC, and the reduction time of the DM \cite{Ferreira2018}.

\subsection{Next-generation adaptive optics control}\label{smarterAO}
This section discusses several possibilities for future directions in RTC for AO.  In present-day conventional AO systems, including the current generation of extreme-AO systems, the RTC follows a well-defined linear control scheme: input measurements from a single WFS are multiplied by a control matrix to produce an output DM control command. Thanks to recent high-performance computing capabilities and new algorithms allowing greater computation speed and power, new extensions to this framework are now emerging to help address various sources of error. 
Below are a few examples: 
\\

\textbf{Sensor fusion techniques} can ingest measurements from multiple WFSs and possibly other sensors in the AO system, aiming to improve the correction's sensitivity, completeness, and robustness. Sensitivity is improved by combining WFSs that, together, lead to a greater resolution of the wavefront. Completeness is improved by gathering information from WFSs that are complementary to the measurements they perform. Robustness gains result from eliminating measurement ambiguities. Sensor fusion techniques, especially those that include  focal plane wavefront sensing, contribute to addressing the non-common path aberrations (NCPAs). \\

\textbf{Predictive control} improves the AO system's temporal error without sacrificing measurement noise. It is the solution to the AO system's most fundamental limit: the trade-off between flux and correction speed. In a conventional AO system, the WFS measurement must be sufficiently fast to keep up with rapidly changing turbulence, but sufficiently slow to average photon noise, and overcome readout noise. The conventional RTC operates at a speed that is a painful compromise between time lag and measurement noise. With predictive control, the time lag is mitigated and replaced by a much smaller temporal prediction error. Recent measurements can be extrapolated in a way that offers the noise-reduction benefits of time averaging without incurring the temporal lag cost \cite{GuyonMales2017}. Several studies on predictive control have recently been published on that topic \cite{VanKooten2020, VanKooten2022, Wong2021, Fowler2022}. \\

\textbf{Non-linear control} extends the range of WFS solutions to include, for example, focal plane images (but not necessarily), in which information about optical aberrations is encoded with high sensitivity,s but for which the relationship to measurements is highly non-linear and induces inaccurate corrections over the wavefront. Artificial Intelligence (AI) plays a significant role in improving non-linear controls, particularly through techniques such as machine learning, neural networks, reinforcement learning \cite{TomeoPou2024RL}, and has been playing a growing role in the last couple of years \cite{Fowler2023}. For example, reinforcement learning allows a system to learn optimal control strategies through trial and error, adjusting to non-linear dynamics by maximizing a reward signal. \\

\textbf{Self-calibration} is essential for the robust operation of the above techniques. In conventional AO, calibration is performed offline prior to running the AO system. Typically, a response matrix will be measured by actuating individual DM actuators or modes and recording the corresponding linear response in the WFS. This approach cannot measure the actual on-sky system response which evolves due to variations in conditions, and it cannot extend to the sensor fusion, predictive control, and non-linear control schemes described above. Instead, the calibration needs to be continuously derived from recent (approximately less than 10 mn for predictive control) on-sky measurements \cite{Skafdrwho, Skaf2022spie}. 

\subsection{The high-performance computing world of RTCs}
\subsubsection{Introduction}
High-performance computing (HPC) refers to the software technique designed to get the most out of the use of computers, often supercomputers. The hardware provides the computing power (whether it is GPU, CPU or FPGA), the HPC uses parallel processing techniques to solve complex computational problems quickly and efficiently. HPC systems typically consist of multiple processors or compute nodes that work together to perform computations in parallel, allowing them to tackle large-scale problems that would be impractical or impossible to solve with traditional computing methods. HPC systems often utilize specialized hardware components optimized for computationally demanding tasks, as for example graphics processing units (GPUs). 
HPC is crucial for RTC systems for several reasons, and has been an active field of development \cite{gratadour2012, Gratadour2018, Guyon2018, Ferreira2020, Ltaief2022}. 
\begin{itemize}
    \item Speed: RTCs require rapid processing of large datasets to make timely corrections. HPC provides the necessary computational power to handle these tasks quickly, ensuring that control actions are executed within the required time frames, often in milliseconds or microseconds. 
    \item Complexity: RTCs involve complex algorithms and computations, which HPC enables by providing the computational resources needed to perform complex calculations in real-time. 
    \item Accuracy: RTCs often require high precision and accuracy in decision-making to ensure the safety and reliability of the controlled processes. HPC facilitates the use of advanced algorithms and numerical techniques that can produce accurate results even under tight time constraints.
    \item Reliability: RTCs must be resilient to failures to ensure uninterrupted operations in the right order, and minimize the impact of hardware or software failures on system performance and reliability. 
\end{itemize}

\subsubsection{How to determine the HPC needs for a RTC system: the example of the Matrix-Vector Multiply (MVM) and GPU/CPU Specs}

The most compute-heavy operation in closing the AO loop is often the matrix-vector-multiply (MVM) converting the input WFS pixel values to output wavefront modes. This MVM must be completed in a fraction of the AO loop period, typically well under 1 ms. 
The MVM computation time is most often limited by the number of memory accesses needed to perform the computation: it is a memory-bound problem. Then, the bigger the memory bandwidth of the hardware is, the faster is the computation time. 
Taking, for example, a large system with 87k input pixels, 33k output modes. Performing this MVM will require to access 11.4 GB of data (considering single precision, i.e. 32 bit per element). Considering the requirement of 1 ms for the RTC time budget, it leads to a requirement of 11.4 TB/s in terms of memory bandwidth.

As of today (year 2024), current GPU such as NVIDIA H100\footnote{More information can be found here: \href{https://www.nvidia.com/en-us/data-center/h100/}{https://www.nvidia.com/en-us/data-center/h100/}} have memory bandwidth of approximately 4 TB/s (note this is terabytes, not terabits). Comparing these specs with the requirement derived above reveals that the system will need at least 3 H100 GPU to do the computation on time.

\textbf{SCExAO computer as an on-sky RTC example:}\\
SCExAO relies on five different computers. The computer on which the CACAO RTC runs is called \texttt{scexaortc}, 
and is the most powerful of all the SCExAO computers, with the most demanding high-performance computing requirements. It has two AMD Epyc 7763 processors with 128 cores total, 1TB RAM, PCI Gen 4, 5 GPUs (A6000, 3080Ti, 2080Ti), and 600TB+ of storage, including 20GB/s+ temp storage SSDs. \texttt{scexao6} is operating a fast 100Gbps downlink from the instrument (located in the IR Nasmyth platform of the Subaru Telescope) to the server room without data collisions and without introducing too much timing variability.
It deals with managing everything linked with the RTC, and controls the following aspects:
\begin{itemize}
    \item the PyWFS data processing to drive the ExAO loop; 
    \item the DM display and control;
    \item the Tip-Tilt (TT) loop;
    \item the astrogrid status;
    \item the low-order WFS loop. 
\end{itemize}

SCExAO performs the second-stage AO thanks to a visible pyramid WFS, operating in the 600-950 nm wavelength range \cite{Lozi2019PWFS}. 
It controls a 2000-actuator MEMS DM from Boston Micromachines. There are 45 actuators across the pupil, giving a control region of $\pm$ 22.5 $\lambda/D$ squared.
The Compute And Control for Adaptive Optic (CACAO) software manages the real-time control (RTC) of the AO system \cite{cacao2018}. 
We use the CACAO software to interact with the system and implement the algorithm to communicate between the PyWFS and the DM.
CACAO enables the control of the DM through 12 different channels, and each channel is a potential input map, offering a layer of correction, given that several corrections are implemented from different wavefront sensors or wavefront control techniques. CACAO then sums up all the maps provided by the user and applies the summed map to the DM. This allows for summing different corrections in parallel, based on multiple WFSs and control algorithms.

\textbf{The SEAL computer as a bench RTC example}:

SEAL \cite{jensenclem2021seal} has an Intel Xeon Gold 6126 CPU operating at 2.60GHz, and an NVIDIA GPU GeForce GTX 1080. SEAL has Python and Julia interfaces to the krtc / DAO real-time control system \cite{Cetre2018}, which is an off-shoot of CACAO. Since SEAL's focus is technology demonstrations, implementations of experiments often prioritize flexibility over speed and therefore do not yet focus on high-speed operations. However, experiments can achieve high AO loop rates when required. An implementation of linear-quadratic-Gaussian control\cite{SenguptaLQG} operated at 100 Hz, with the main limiting factor being the difficulty of synchronization between the WFS, DM, and simulated disturbances across threads. This was a limitation imposed by the Python interface rather than the hardware or RTC, and it was addressed within these constraints by spin-locking to ensure precise timing across threads. Similarly, tests of multi-WFS single-conjugate AO\cite{Gerardmultiwfs} successfully operated with two AO loops in different Python sessions at 100 Hz and 200 Hz. Most of the latency in these cases came from the reconstruction algorithm for the FAST focal-plane wavefront sensor. The algorithm required two FFTs, requiring 5ms with significant jitter on scales that affected the AO loop rate. Using quicker implementations, for example via the Julia interface and/or GPU, or computationally easier AO algorithms, could enable even faster operation up to kHz scales.

SEAL is also often affected by latency due to concurrent hardware access. Operating several cameras at once can limit the loop rate of each one, and in some cases can completely stall operations. Updates to the RTC or the interface may help to address this.

\section{Shared-memories in RTC}\label{sec:shm}

\subsection{Shared memory data management description}

Shared memory refers to a type of inter-process communication (IPC) mechanism in computing where multiple processes can access common memory locations. In shared memory systems, processes can communicate and synchronize by reading from and writing to the same memory addresses. This shared memory region is typically managed by the operating system and can be accessed by multiple processes concurrently. Shared memory is a fundamental concept in parallel and distributed computing, enabling efficient communication and coordination among processes or threads sharing a common address space. It is widely used in various computing environments, and ideally suited for RTC. It is especially an efficient data management method, because it avoids the overhead associated with copying data between processes. Since processes directly access shared memory locations, data can be transferred quickly and with low latency. Shared memory requires mechanisms for synchronizing access to shared data to prevent race conditions and ensure consistency. Techniques such as semaphores are commonly used to coordinate access to shared resources and enforce mutual exclusion among processes.

\subsection{Problem statement and intention}

The realm of RTC systems is software-centric, posing significant challenges for astronomers and WFS\&C scientists who often lack the specialized training needed to directly modify RTC code. Developing a thorough understanding of the RTC's behavior, parameters, and the impact of various inputs on outputs can be the work of months or years, depending on the background in software engineering. Moreover, the interfaces for RTC systems often present complex information, leading to information overload on the user end.
To address some of these challenges, some RTC developers across various AO systems have adopted a standard for data management and interaction: the shared-memory architecture. This section describes the advantages of shared-memory architecture and its importance for facilitating ease of use and collaboration in RTC development. By using a shared-memory approach, and allowing access to the shared-memories, the interaction with the RTC becomes more transparent, allowing users to implement their algorithms without needing to understand the details of the system's inner workings. 

\begin{figure}[ht]
\begin{center}
\includegraphics[width=0.7\textwidth]{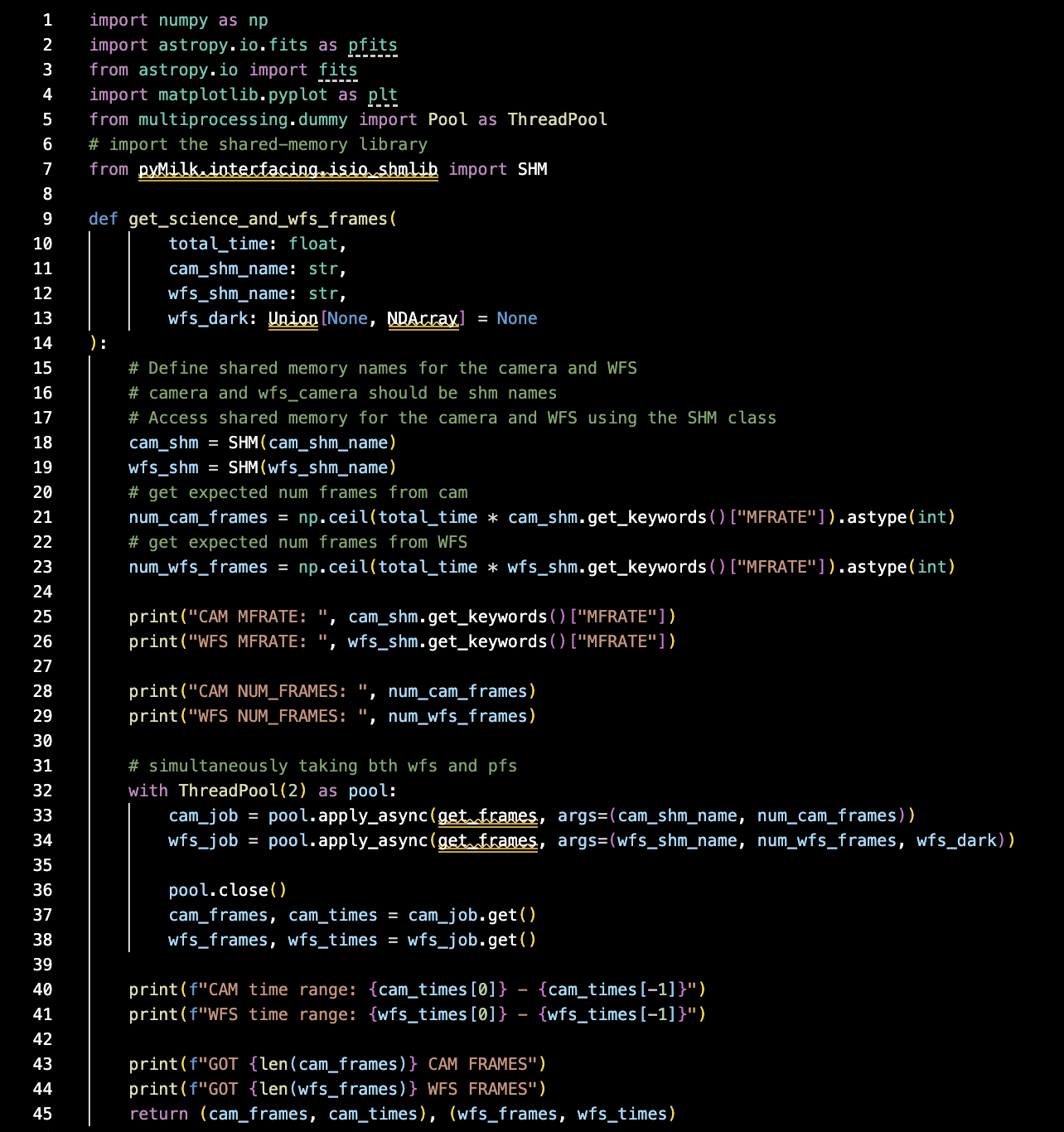}
\caption{%
 Example of a python function that gets PSF and WFS frames and synchronizes them. Lines 11 and 12 correspond to the name associated to the shared memory of the PSF and WFS respectively (for example, "pueo" and "vcam1" for SCExAO's pyramid WFS and Vampires' first camera). These two lines are sufficient alone, the rest is simply an example of use. Lines 18 and 19 then assign a name for these specific shared memories that will be used later in the code. In this case, it is to allow for changing which shared memory one wants to work with. Line 21 and 23 shows an example of a function that gives some information on the shared memory, in this case the keyword for each frame. } 
\label{fig:shm_example}
\end{center}
\end{figure}

The user still needs to understand how the data in the shared memory is treated by the RTC, hence there must be an understanding of how the wavefront control happens within the RTC. Ideally, and the RTC community has been working on this aspect, the user does not need to dive into the deeper parts of the software that can remain a form of "black box". Below is a concrete example of a code that takes both the PSF and WFS data from the shared memory, and synchronizes them. Figure \ref{fig:shm_example} shows an example of a script that collects PSF and WFS frames, and synchronizes them.

This architecture not only simplifies the interface for end-users but also fosters collaborative work by providing a common ground for data interaction. It enables closer-to-seamless integration and communication between different components of the system, promoting a cooperative environment where complex tasks are made more manageable. The shared-memory system thus offers an elegant solution to the inherent challenges of RTC systems, making the technology more accessible and inspiring collaborative efforts in the pursuit of scientific progress. In some cases, however, the shared-memories are nested with the real-time processes, which is necessary to understand to keep an optimal performance. In such a case, providing free access to all the shared memories to the user can bring conflicts within the processes, and some communication with the RTC developers is necessary to avoid any issues.


\subsection{The CACAO RTC}\label{sec:cacao}
CACAO\footnote{Open source, available here: \href{https://github.com/cacao-org/cacao} (Compute And Control for Adaptive Optics), initially designed for the SCExAO instrument at the Subaru Telescope, is a mature, flexible, modular and open-source software. It is designed to benefit from any available CPUs and GPUs, and to manage data flows from various sources (wavefront sensors, cameras, other telemetry) and with different frame rates. CACAO is mainly coded in C, which provides fast execution speed, making it the language of choice for real-time algorithms.  \\
A thorough description of CACAO {\texttt{https://github.com/cacao-org/cacao}}.} is presented in several proceedings \cite{cacao2018} \cite{Guyoncacao2020}, and in the proceeding of Deo et al. 2024. Because many RTCs worldwide are based on some aspects of CACAO (mainly the shared-memory structure), or got inspired by it, we present here a description of the CACAO RTC, its architecture, and its data management system. This is targeted to  WFS\&C users and astronomers rather than RTC developers.

\subsubsection{CACAO data management}

\begin{figure}[t]
\begin{center}
\includegraphics[width=0.9\textwidth]{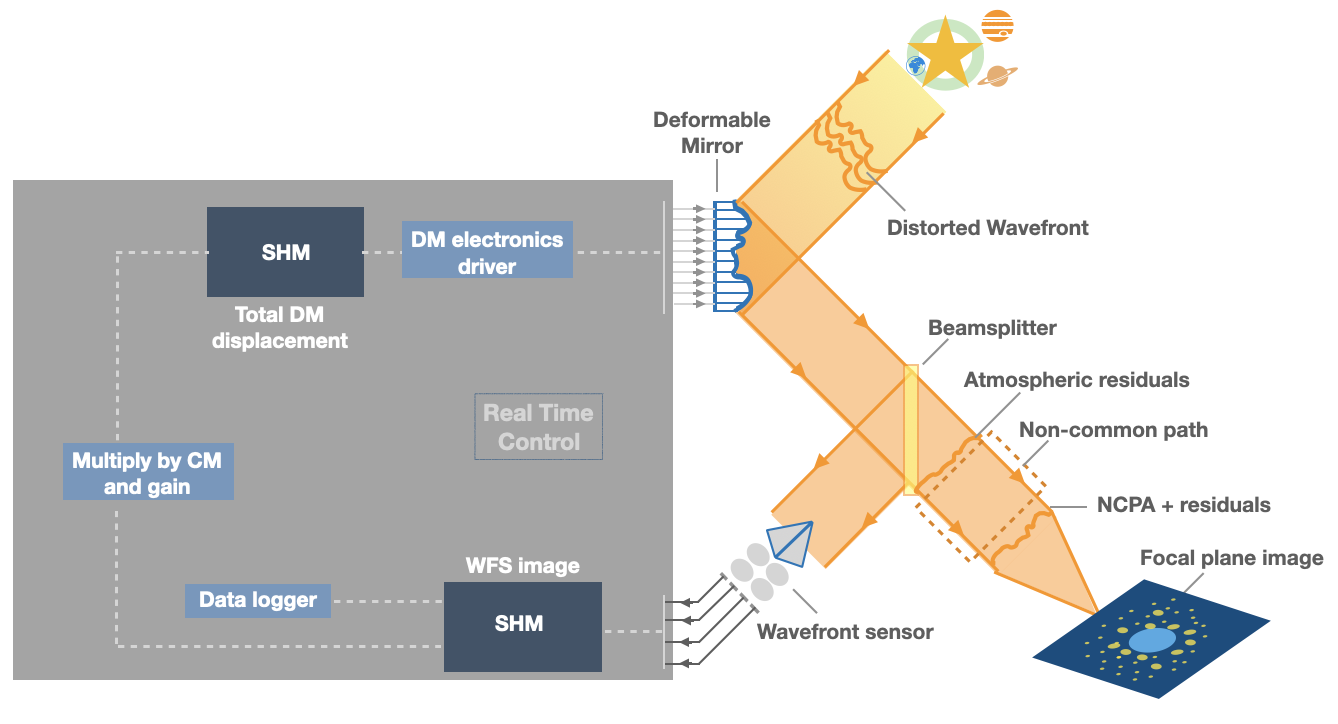}
\caption{%
Schematic view of the data management with shared memories between the WFS and the DM. The light blue boxes are processes, linked together with streams, represented in dark blue boxes. The signaling between streams and processes is achieved through semaphores: a process "waits" to a semaphore, and will start as soon as the semaphore is "posted" to 1, at which point it will decrement the semaphore value back to zero. Figure adapted from \cite{Skafdrwho}.}
\label{fig:data_management}
\end{center}
\end{figure}

The data flow from the different streams is managed with  shared memory structures, accessible directly from computer RAM. 
To avoid any collision, a system of semaphores, locking shared memory buffers, arises when waiting for the appropriate timing for the right data flow. There are ten stream semaphores. This architecture allows users to plug and play their own codes without diving into the hardware camera interfaces. 
Figure \ref{fig:data_management} presents a schematic view of this shared memory architecture, and the steps taken within the RTC in an AO system. \\

\begin{figure}[t]
\begin{center}
\includegraphics[width=1\textwidth]{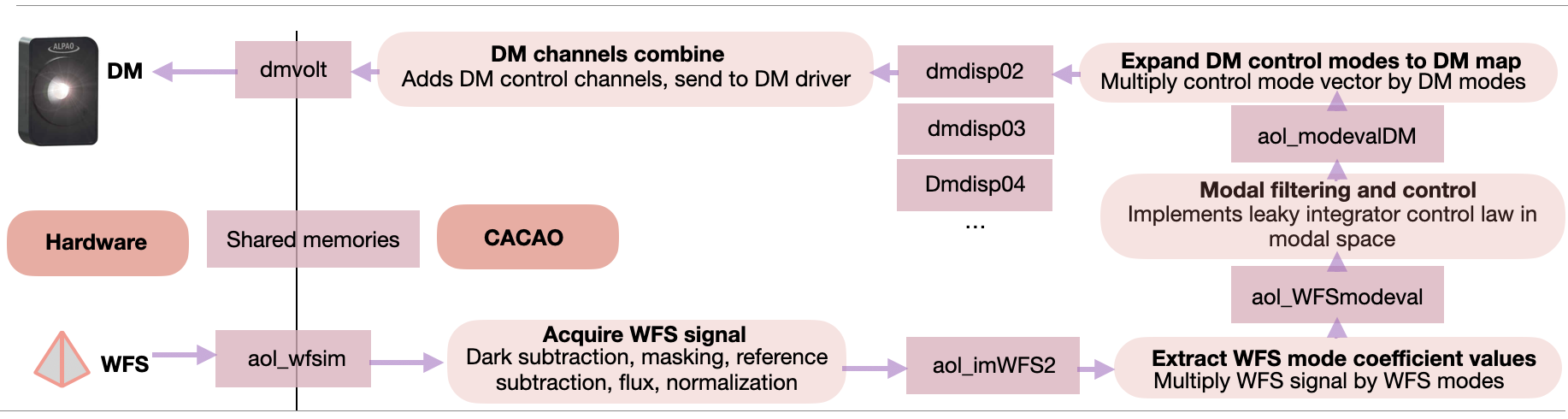}
\caption{%
Canonical modal control in CACAO. The pink rectangular boxes represent shared memories. CACAO first reads the shared-memory that is given by the WFS, canonically called ``aol\_wfsim''. It then processes the signal and writes into the shared-memory named ``dmvolt'', which is then read by the DM. }
\label{fig:cacao_canonic}
\end{center}
\end{figure}

\textbf{ImageStreamIO}\\
\texttt{ImageStreamIO}\footnote{Open source, available here: \href{https://github.com/milk-org/ImageStreamIO}{\texttt{https://github.com/milk-org/ImageStreamIO}}.} is the core library of all of CACAO: it contains all the data format and synchronization.
It allows the use of the data streams in shared memory image format. This deals directly with the camera FITS files, allowing to read and write data streams, at low latency and a high throughput input/output (I/O) of data. There are two main scripts for data conversion, either converting from shared memory to FITS files (\texttt{shmim2fits}) or converting from FITS files to shared memory (\texttt{Fits2shm}). \\

\textbf{Milk}\\
The \texttt{Milk} (Modular Image processing Library toolKit) software framework\footnote{Open source, available here: \href{https://github.com/milk-org/milk}{\texttt{https://github.com/milk-org/milk}}.} mainly manages the data architecture, providing tools for image processing and analysis in real-time. \texttt{Milk} also provides a command line interface for easy access to functions and arguments, manages runtime loading of modules to extend capabilities, and provides inter-process communication (IPC), for sharing process status and function parameters, and images. \texttt{Milk} is coded in C, which isn't the most used language in the high-contrast imaging community. To make it more user-friendly, a python version was developed, \texttt{pyMilk}\footnote{Open source, available here: \href{https://github.com/milk-org/pyMilk}{\texttt{https://github.com/milk-org/pyMilk}}.}, which is a python wrapper to ImageStreamIO. It provides an easy way to link python scripts and the shared memory structures used in \texttt{Milk}. 

\subsection{Hardware interface}\label{sec:hardware}
The hardware interface is often the main challenge when interfacing with a software. 
Connecting with the hardware drivers requires working with the libraries of the company of the hardware. Each hardware company has their own software that needs to be communicated with to translate the data into shared memory format. 

Understanding the various hardware layers involved in communication is essential for effective integration and operation of systems like cameras. The communication stack of a given camera typically includes multiple layers, each with specific functions, including:

\begin{itemize}
    \item Physical layer: this is the most fundamental layer, consisting of the camera's physical components and the connections to the frame grabber. The physical layer involves the conversion of light into electrical signals by the camera's sensor.
    \item Frame Grabber: this is a crucial hardware component that captures the electrical signals from the camera's sensor. It contains electronic chips and associated firmware. The firmware provides instructions to the chips, enabling them to convert the electrical signals (voltages) into digital pixel data.
    \item Driver: the driver is a software component that interfaces with the operating system (OS). It translates the pixel data from the frame grabber into a format that the OS can understand and process. The driver ensures that the OS can read the pixel data and make it accessible to other software applications on the computer.
    \item Application Programming Interface (API): the API is a software layer that provides a set of routines and protocols for interacting with the hardware (e.g., the electronic card within the frame grabber). The API abstracts the hardware details and offers a simplified interface for users to use in their applications.
    \item Software Development Kit (SDK): the SDK is an extension of the API, providing additional tools, libraries, and documentation to facilitate software development. The SDK includes a user interface and sample code, making it easier for users to create applications that interact with the camera and process the images. It is through the SDK that the user communicates with everything else. 
\end{itemize}

Effective hardware communication and integration requires a clear understanding of these layers. Each layer plays a critical role in converting raw sensor data into usable digital information, ultimately enabling the development of sophisticated applications. Usually for the user, the word "driver" is often used to refer to all of these elements. 



\begin{figure}[t]
\begin{center}
\includegraphics[width=0.5\textwidth]{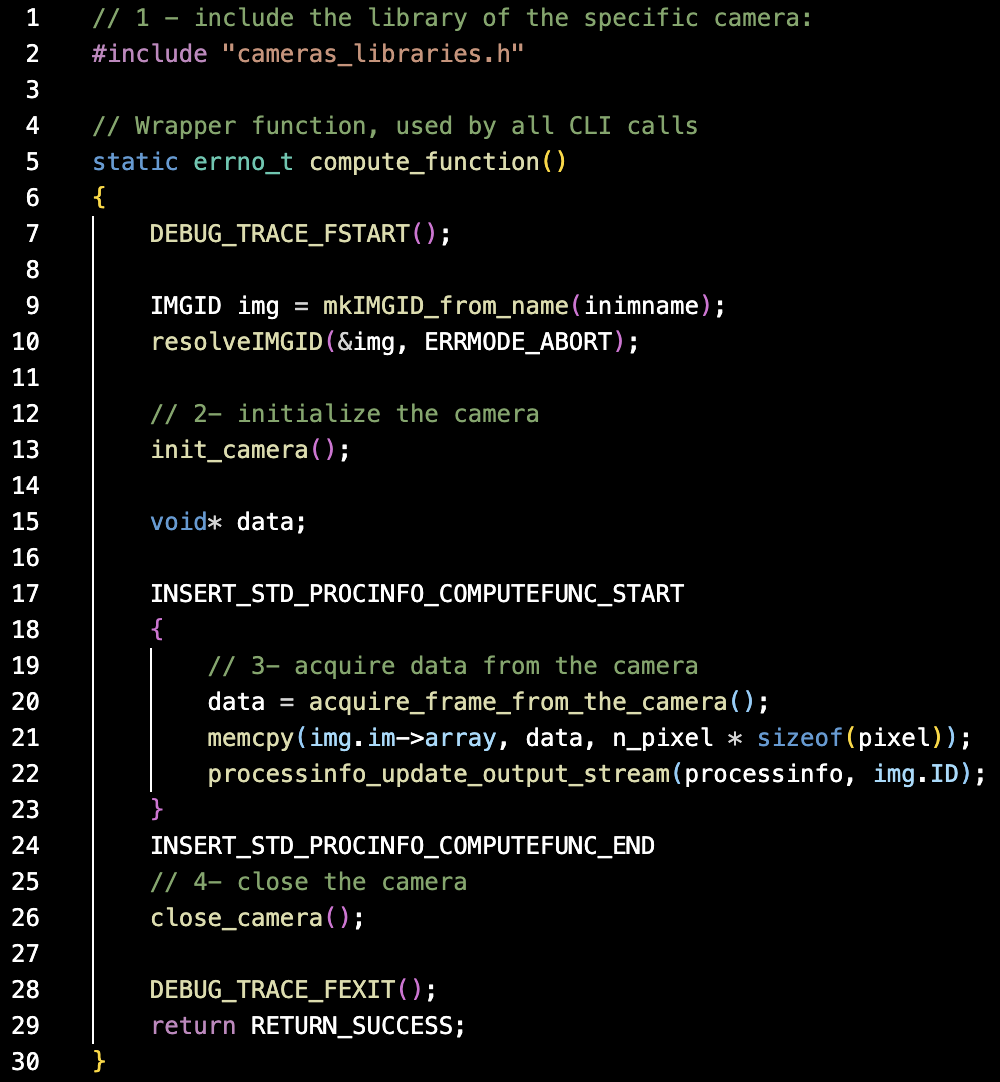}
\caption{%
Example of the creation of a driver to use a camera library to turn the data into a shared-memory stream.
The code is available at: 
\href{https://github.com/milk-org/milk/blob/dev/src/milk_module_example/examplefunc3_updatestreamloop.c}{\small \texttt{https://github.com/milk-org/milk/blob/dev/src/milk\_module\_example/examplefunc3\_updatestreamloop.c}}}
\label{fig:camera_to_shm}
\end{center}
\end{figure}

CACAO has adopted a straightforward approach to interfacing custom software with dedicated hardware, where one stream is defined as a hardware endpoint and is the only interface exposed to the rest of the CACAO ecosystem. SCExAO has developed a collection of interfaces that cover a variety of scientific camera manufacturers (Andor, First Light Imaging, Teledyne/FLIR, Hamamatsu, Nuvu, ...) as well as deformable mirrors (Boston Micromachines, ALPAO). 
The easiest hardware integration is with hardware that was already integrated into the shared-memory format with the manufacturers mentioned above. If this is not the case, the steps for successful hardware integration are to first work with the hardware manufacturer and have the library run in C or C++. Then, this can be used in a CACAO template that will transfer the data into a shared-memory. The template can be found in the Milk module, Figure \ref{fig:camera_to_shm} presents the four main steps to create a new driver from scratch. 

\section{Rational for shared-memory-based RTC}\label{sec:shmrational}
This section highlights the benefits of using the shared-memory solution for data management. Although the shared-memory isn't the solution to all the problems, the main point here is to present this tool as a form of standard used by some of the RTC community for increased collaboration. 

\subsection{Highlight of flexibility and cost}

The use of shared-memory based RTC systems introduces significant flexibility in hardware selection, allowing for the use of "commercial off-the-shelf" (COTS) components. This approach highlights the primary advantage of shared-memory systems: hardware agnosticism. The software operates independently of the underlying hardware, enabling nearly seamless hardware upgrades with minimal software modifications and simplifying the integration of spare parts.

This flexibility offers several benefits. Firstly, new developments, including the integration of new hardware from familiar manufacturers, can be expedited. Secondly, redeploying hardware from one computer to another becomes more easily achievable, as reconfiguration is relatively rather straightforward. Lastly, the relative simplicity of the hardware-software interface, compared to more complex software stacks, allows users to focus on performance optimization rather than extensive feature coverage.

\subsection{Collaborative development and open source}
Whether a specific RTC is open source or close source, the use of shared-memories allows for greater collaborative opportunities through efficient data sharing, and hence seamless integration of hardware and WFS\&C algorithms. Shared memories furthermore provides a standardized method of communication with the different hardware components in different AO systems. This standardization is crucial for collaboration, as it enables communication with hardware components from different manufacturers or research groups to interoperate smoothly, facilitating projects that involve several institutions with common endeavors. Indeed, several instruments and benches worldwide use shared-memory-based RTC, from CACAO's ImageStreamIO format. Below is a non-exhaustive list of these cases, we will expand in the next session on some of these RTCs.

CACAO was adopted for the control of the MagAO-X \cite{Males2018MagAOx} platform
at the Magellan telescope in Chile. CACAO is the RTC solution used by the MAPS project \cite{MAPS_Morzinski_2020}, currently mid-commissioning at the 6.5-m MMT telescope in Arizona. CACAO is used by the experimental platform KalAO \cite{Hagelberg_janis_kalAO_2020}, led by the observatory of Geneva and operational at the 1.2 m Euler Telescope at La Silla Observatory. Promoted by these early adopters, CACAO is already the chosen solution for future projects at Subaru Telescope — the AO3K [59] upgrade and the ULTIMATE laser tomography [60,61] system; and for the GMagAO-X high-contrast platform on the upcoming 24 m Giant Magellan telescope [19]. CACAO's shared-memory format is also at the base of the COSMIC RTC, which is the RTC for MICADO on the E-ELT \cite{FerreiraCosmic2020}, Keck I and II \cite{Chin2022_rtc_keck}, as well as MAVIS on the VLT \cite{Bernard2024}. Likewise, the DAO RTC \cite{Cetre2018}, which is the RTC for  HARMONI on the E-ELT.
Along with these instruments, some optical benches, such as HiCAT at STSCI, got inspired by CACAO's shared-memory data management style \cite{por2024catkit2}.



\subsection{A few shm-based RTC examples}\label{sec:diffRTC}

\subsubsection{MagAO-X} 
The Magellan Adaptive Optics Extreme - MagAO-X - instrument,  is installed on the 6.5 m Magellan Clay telescope at Las Campanas Observatory in Chile. It is the prototype platform for the GMagAO-X instrument, which will be on the Giant Magellan Telescope (GMT) \cite{Males2024gmagaox}. MagAO-X is designed for HCI of exoplanetary systems.  Similar to SCExAO, MagAO-X has a dual purpose: WSF\&C development, along with astronomy-science observations.  

MagAO-X uses CACAO for its core RTC functionality.  The MagAO-X instrument control software system is a c++ application framework designed to work with CACAO.  In particular, it uses the ImageStreamIO library for all image acquisition, processing, and low-latency IPC.  A key rationale for this architecture is that any camera, including science detectors, can be used as a wavefront sensor feeding corrections to any deformable mirror (or other corrector, such as tip/tilt stages).  

A unique feature of MagAO-X is that it uses a woofer-tweeter configuration, using offloading, It uses a woofer-tweeter configuration to provide sufficient low-order stroke while enabling high order correction with a MEMS DM. The shape on the tweeter is averaged for a configurable number of loop steps. That average shape is then multiplied by the tweeter-to-woofer response matrix, and the result is applied to the woofer with a gain and a leak.  With long averaging ($\sim$10 loop steps) and a small gain it causes a small perturbation to the high-order loop.  The tweeter-to-woofer response matrix is calibrated from the woofer and tweeter zonal response matrices taken on the WFS in the same alignment.  The offloading modes are normally filtered, with typically only 5 to 10 low-order modes being offloaded (which depends on the observing conditions). \\

\subsubsection{MAPS}

The MMT AO exoPlanet characterization System (MAPS) is an extensive upgrade to the adaptive secondary mirror (ASM) AO system at the 6.5 m MMT telescope on Mt. Hopkins, Arizona\cite{MAPS_Morzinski_2020} (see Morzinski et al, in these proceedings).  The new MAPS system includes both visible and IR pyramid sensors, as well as several IR science instruments. Like MagAO-X, MAPS uses CACAO for its core RTC functionality and MILK and ImageStreamIO are used extensively for instrument control.  MAPS has two interesting features compared to typical CACAO-based systems: it uses pyramid slopes, and the 336 actuators of the ASM are not laid out in a regular grid.  Slopes are calculated by the instrument software, and the output is treated as the raw WFS signal by CACAO. The use of slopes necessitates several differences in the pre-processing done by CACAO, e.g. no dark or reference subtraction. The DM command is handled as a 1x336 vector.  Tools were implemented to enable 2D display of these commands.

\subsubsection{The DAO RTC}\label{sec:DAO}
The Durham Adaptive Optics (DAO) Real-Time Controller (RTC) is designed for the HARMONI and MOSAIC instruments on the E-ELT \cite{Harmoni_2022}. It utilizes a shared-memory format originally forked from ImageStreamIO in 2016, following work on the IR pyramid RTC at W.M. Keck Observatory. Until recently, DAO was still based on this fork, but the team has since offered an option to resynchronize the ImageStreamIO component, making DAO fully compatible with other CACAO RTC systems. DAO primarily operates on a CPU but also provides GPU solutions. It offers a suite of tools dedicated to AO and supports a wide range of hardware.
Although developed mainly for HARMONI, DAO is now used in other systems at Santa Cruz on SEAL for certain cameras. A collaboration has begun with the INAF Padova Observatory to build their new MCAO bench. A new DAO RTC has been deployed on PAPYRUS at the Observatoire de Haute Provence (OHP) in France. DAO will also be used for the European Solar Telescope and a turbulence profiler for DKIST. Additionally, DAO is used at Durham University for multiple AO projects (WIVERN, NESSIE, MKIDS) and is a potential solution for the VLT RISTRETTO instrument. 

\subsubsection{The COSMIC RTC}
COSMIC is a robust software stack built around the shared-memory paradigm, providing a stable high-level API to its users. COSMIC aims to abstract away all low-level interface and function calls, allowing the user/AO scientist to configure a set of processes in high-level configuration files. This \textit{turn-key} solution has been proven on the Keck telescopes, and is proposed for a number of other instruments with even higher demand (e.g., HARMONI, MAVIS\cite{Bernard2024,Gratadour2022}). 

COSMIC is fast and scalable (see, e.g., \citenum{FerreiraCosmic2020}), but historically has been closed-source, introducing a challenge to users wishing to deviate from classical RTC functionality. That said, an open-source release of COSMIC has recently (June, 2024) been announced. Such a release would improve accessibility for a host of research institutions who wish to develop novel RTC techniques for themselves and the community.

\subsubsection{ImageStreamIO in space} 
WFS\&C for space-based telescopes is fundamentally different from that for ground-based systems. Unlike ground-based telescopes, space telescopes do not need to contend with real-time atmospheric turbulence corrections. Instead, the aberrations in space-based systems are typically induced by the instrument's optics and mechanical and thermal disturbances from the telescope and spacecraft. Moreover, space-based wavefront control often focuses on optimizing the wavefront to create a dark hole, thereby increasing the contrast necessary for direct imaging of exoplanets.
Using shared-memory architectures for WFS\&C in space telescopes offers several significant advantages. Firstly, having a control system onboard the spacecraft provides a more cost-effective and relevant means of managing the wavefront. This onboard control system reduces the latency and complexity associated with real-time adjustments from ground-based operations. Additionally, it facilitates seamless collaboration with ground-based telescopes, allowing for the integration of advanced WFS\&C techniques developed for terrestrial applications.

A prime example of this approach is the Space Coronagraph Optical Bench (SCoOB)\cite{ashcraft_space_2022} which is demonstrating flight ready technologies wavefront sensing and control for potential future missions such as the Coronagraphic Debris and Exoplanet Exploring Pioneer concept \cite{maier_design_2020} and other low-cost exoplanet imaging missions under study \cite{Douglas2023}. SCoOB is equipped with a high-contrast ($<$1e-8, Van Gorkom et al, these proceedings) coronagraph and utilizes a wavefront control system based on a shared-memory architecture derived from CACAO and MagAO-X\footnote{Here is the GitHub page: \href{https://github.com/uasal/MagAOX-scoob}{\texttt{https://github.com/uasal/MagAOX-scoob}}}. This shared-memory framework enhances the efficiency and flexibility of the control system, allowing for smoother software updates compared to how other spacecrafts are dealing with their control computers. It also supports collaborative efforts by enabling the reuse of established algorithms and techniques from ground-based AO systems, including leveraging space-rated GPUs \cite{GPU4space2021,Belsten2023}. 

The implementation of shared-memory architectures in space-based WFS\&C systems, such as SCoOB, underscores the potential for combining and using WFS\&C systems from both space-based and ground-based telescopes, leveraging shared-memory architectures to enhance overall performance and collaboration. This integration can lead to more efficient and advanced technological solutions across different types of astronomical instruments and environments.

\subsubsection{The HiCAT RTC}

The High-Contrast Imager for Complex Aperture Telescopes (HiCAT) testbed\cite{SoummerSPIE2024}, located at STScI, is a testbed dedicated to system-level technology demonstration of high-contrast imaging on segmented telescopes. With future space telescopes requiring multiple concurrent control loops to stabilize the wavefront, HiCAT transitioned to a shared-memory-based RTC software, called catkit2\cite{por2024catkit2}, running on off-the-shelf computing hardware. For hardware-compatibility reasons, catkit2 uses their own shared-memory format inspired by ImageStreamIO. In addition, catkit2 uses a service-oriented architecture to manage all RTC processes and data streams. This process management, and accompanying service discovery, enables increased hardware safety for performing long-running experiments without human intervention. Being written in C++ and Python, it enables fast prototyping of new wavefront control algorithms \cite{redmond2022maintenance_hicat,will2023adwc_experimental} while maintaining easy integration of new hardware.

Although developed for the HiCAT testbed, catkit2 is starting to be used at other testbeds. The Tr\`{e}s Haute Dynamique 2 (THD2) testbed located at the Observatoire de Paris, is currently transitioning to catkit2 as their new RTC.

\section{SEAL Testbed at UCSC: RTC Upgrade}
\subsection{Introduction to the SEAL testbed}
The Santa Cruz Extreme AO Lab (SEAL)\cite{jensenclem2021seal} is a testbed in the visible and infrared, designed to advance the state of the art in wavefront control for high contrast imaging on large, segmented, ground-based telescopes. 

\begin{figure}[t]
\begin{center}
\includegraphics[width=0.9\textwidth]{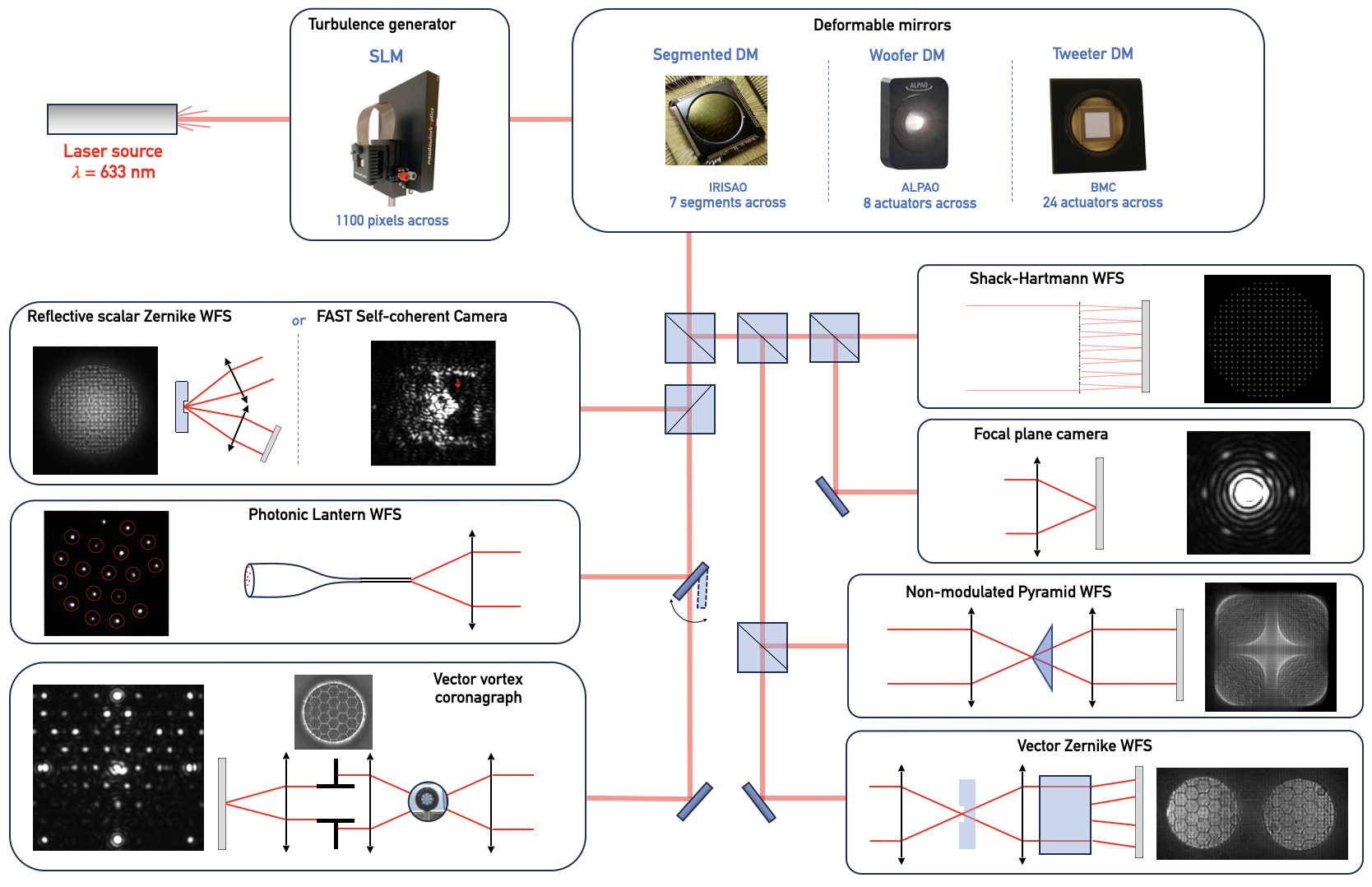}
\caption{%
Schematic layout of the SEAL testbed.}
\label{fig:seal_layout}
\end{center}
\end{figure}

Figure \ref{fig:seal_layout} presents the schematic optical and hardware layout of the SEAL testbed. The AO system comprises a pair of deformable mirrors, with one acting as a woofer with 97 actuators from ALPAO DM and the other as a tweeter with 1024 actuators from Boston Micromachines MEMs DM. Additionally, it includes four wavefront sensor arms: 
\begin{itemize}
    \item a high-speed Shack-Hartmann WFS,
    \item a reflective pyramid WFS,
    \item a vector-Zernike WFS \cite{Salama2022} ,
    \item a self-coherent camera \cite{Gerard2021}, 
\end{itemize}

The testbed also features science arms initially equipped with a vector-vortex coronagraph and the self-coherent camera, which serves both as a wavefront sensor and a science camera. 
SEAL's initial RTC system, named kRTC, is an off-shoot of the CACAO RTC, and featured several differences. The intention was to facilitate the seamless exchange of software between SEAL and the Keck II AO system, given kRTC was the implemented RTC on the infrared pyramid WFS.

\subsection{RTC challenges and requirements}
AO benches and instruments on telescopes face different challenges and requirements. Benches have a more controlled and stable environment, with easy accessibility for adjustments, maintenance and upgrade, while not having the responsibility to perform observations on sky as instruments on telescopes do. That main difference leads to a high flexibility and modularity that makes benches ideal for performing WFS\&C development, with modifying, adding and removing hardware components and software upgrades, allowing tests of new technologies and refine algorithms before deploying them on a telescope. Furthermore, AO benches do not have the same real-time needs as instruments do, hence the latency can be very slow. 
In the context of SEAL, the long-term intention is to implement new WFS\&C technologies on the Keck telescope. In 2024, Keck's pyramid wavefront sensor and associated kRTC machine were retired, and the Keck II AO system RTC was upgraded to the COSMIC RTC framework. In the future, a development-only CACAO-based RTC at Keck II would allow for technology transfer from other CACAO-bases systems including SEAL in a way that does not affect the operational COSMIC-based system.

\subsection{Implementing CACAO on SEAL}
\subsubsection{Current status}
In any RTC implementation, the communication with the hardware remains the main challenge, as described in Section \ref{sec:hardware}. With kRTC being previously implemented with the SEAL hardware with a different shared-memory format than the one used by CACAO, one of the major benefits in implementing CACAO on SEAL thus far has been to make the shared-memory format compatible between kRTC / DAO and CACAO, as briefly described in \ref{sec:DAO}. Once the shared-memory data structure is implemented on the hardware (WFS and DM) as described in Section \ref{sec:hardware}, there are three main steps for calibration and closing the loop: 
\begin{itemize}
    \item accurately measure the hardware latency (HL);
    \item take a response matrix;
    \item compute a control matrix and close the loop. 
\end{itemize}


The HL is the time delay between issuing a DM command and having the corresponding wavefront change visible in the arriving WFS data. HL includes data transfers between the RTC and hardware devices, digital-to-analog conversion of DM commands, mechanical motion of the DM surface, and WFS camera readout. Measuring HL requires real-time performance, as DM commands must be issued at known time offsets relative to WFS frames arrival. It is implemented in CACAO, and is a powerful diagnostic of all critical hardware as well as RTC configuration.

Accurate knowledge of HL is essential to modern AO systems. Operating an AO loop critically relies on a precise linear mapping between DM commands and WFS signals. Historically, early AO systems measured this response by sequentially poking the DM and waiting sufficient time for the WFS to include the corresponding signal, an approach that does not scale to modern high-order AO systems with thousands of actuators. Instead, a rapid fire of DM pokes should be issued, precisely timed such that the DM moves between $\sim$ kHz WFS frames, allowing a high precision measurement of the system linear response to many actuators.

The step-by-step guide can be found on the CACAO GitHub page: \\
\href{https://github.com/cacao-org/cacao/blob/dev/AOloopControl/examples/scexao-vispyr-bin2-conf/README.md}{\small \texttt{https://github.com/cacao-org/cacao/blob/dev/AOloopControl/examples/scexao-vispyr-bin2-conf/README.md}}

There have been several challenges along the way making the progress tedious. First, the difference of shared memory format was not initially not well understood, it took several iterations to communicate properly with the hardware, both from hardware to software, and from software to hardware. Then, the speed of the pyramid wavefront sensor camera, running at $\sim$ 15Hz: CACAO's HL measurement is not well-suited for such slow speed, making it challenging to map the WFS and DM. The Shack-Hartmann WFS has a greater speed and is more suitable in this specific context.

\subsubsection{Control on SEAL}
SEAL offers a woofer-tweeter configuration. To deal with this configuration, one control method used on the testbed uses the framework developed in Calvin et al. (in prep) for simultaneously controlling a segmented primary mirror along side the AO system, which shares the same mathematics as the modal decoupling algorithm first presented by Conan et al. (2007) \cite{Conan2007woofertweeter}. The interaction matrices for both deformable mirrors are calculated independently, and then the command matrix (found as the pseudo inverse of the interaction matrix) of the primary deformable mirror (in our case, the woofer, for its larger amplitude range) is used to subtract the primary DM's controllable space from the auxiliary's interaction matrix.
Mathematically, this is seen as $IM2\_i = IM2 - IM1 \times IM1^{-1} \times IM2$, where the $i$ represents that the resulting interaction matrix is the "independent" interaction matrix. With this, the pseudo-inverse of the independent IM can identify the signal from the wavefront sensor that cannot be controlled by the woofer and commands the tweeter to correct that independent signal, while the same WFS output simultaneously uses the woofer to correct the wavefront errors within the woofer's controllable space.

\subsection{Status of RTC upgrade on UC Lick Shane and Automated Planet Finder Telescope}

The University of California Lick Observatory, located on Mount Hamilton, has two mid-sized telescopes that are suitable for adopting the CACAO RTC package for their current and future AO systems. 

The Shane telescope has a 3-meter primary mirror and is equipped with an AO system, ShaneAO, that has gone through several generations of upgrades \cite{olivier1995shane, kupke2012shane, gavel2016shane}. With the intention of using Shane as a development platform for the Keck Telescopes, there is an on-going effort in upgrading the RTC of the AO system. Given the telescope is being used for science observations, the current work is focused on installing CACAO on an upgraded computer, that works in parallel with the existing instrument. \\

The Automated Planet Finder (APF) is a 2.4-meter telescope that operates fully robotically to collect precision spectra to discover exoplanets using the radial velocity method. 
A concept for an adaptive secondary mirror based AO system has been proposed for APF \cite{Bowens-Rubin2022} which could more than double the spectrographic throughput. The upgrade would incorporate a  TNO 61-actuator large-format convex deformable  mirror in the position of the secondary mirror.  While the APF adaptive secondary mirror awaits funding, members of the UCSC Lab for Adaptive Optics are building a large-format deformable mirror on-sky testbench that will be located post-focal plane at the open Nasmyth port of the APF. CACAO is currently being developed as the RTC software for this large-format deformable testbench to work in conjunction with the TNO DM3\cite{Bowens-Rubin2020} 57-actuator flat deformable mirror lab prototype.  The software developed as part of these efforts at the APF provides a future pathway to use the CACAO RTC package to control the TNO large-format deformable mirrors when being used as adaptive secondary mirrors. 


\section{Lessons Learned and Best Practices}

\subsection{Collaborative software practices}\label{sec:coll}
To ensure the stability and reproductibility of the AO system configuration, it is imperative to maintain a workflow that documents the optimal setup. It is also important to always secure a working and robust configuration, as updates (e.g., via a \verb|git pull|) may introduce bugs that could disrupt the system functionality. 
In the context of collaborative development across various instruments, effective version control is crucial. It is essential to track the specific Git hash (version number) of functional configurations. This practice enables the identification and replication of stable system states.
To facilitate collaboration and maintain system integrity, each collaborator should maintain their own fork of the repository, periodically synchronizing with the development branch. This approach accounts for the diversity in hardware, operating systems, and drivers (e.g., different NVIDIA driver versions), which necessitates rigorous version control.
In summary, diligent version control and documentation of functional configurations are fundamental to successful collaborative development in AO systems, ensuring compatibility and stability across diverse environments.

\subsection{Successful bench and instrument integration}

Integrating the RTC within an instrument and telescope requires three main considerations.

First, the cyber infrastructure of the telescope has to communicate with the RTC, which requires a multifaceted approach to ensure seamless operation and user efficiency. Firstly, the RTC must be harmoniously integrated with the existing observatory systems, ensuring compatibility and communication between various software and hardware components. This necessitates a comprehensive understanding of the observatory's cyber infrastructure to facilitate smooth data flow and coordination, as each telescope has their own software infrastructure.

Then, the user experience is another critical factor. A well-designed graphical user interface (GUI) is essential to provide intuitive and efficient interaction with the RTC system. The GUI should present complex data and control options in a user-friendly manner, reducing the cognitive load on astronomers and operators. This is particularly important given the intricate nature of AO systems and the real-time adjustments required to correct atmospheric distortions.

Finally, comparable to what is described in Section \ref{sec:coll}, the successful feature integration of the RTC within the telescopes includes implementing robust version control to track changes and maintain a stable system configuration. Each new configuration involves rigorous testing and validation to ensure that new functionalities work seamlessly with existing components.

\section{RTC carbon footprint} 

The carbon footprint of RTC systems and HPC in general is a growing concern as these systems require significant computational power and energy consumption \cite{HPC_astro_co2_2020}. Here's a closer look at the factors that contribute to the carbon footprint of RTC: 
\begin{itemize}
    \item Energy consumption: whether it is from the computational power or the servers cooling systems, RTCs lead to high energy consumption. 
    \item Data management: an average night on the SCExAO instrument at the Subaru telescope corresponds to $\sim$ 10 Tbs of data. Storing this data is not only energy-intensive, but also not sustainable on the long run. 
    \item Hardware: the production of the hardware used in RTC systems, including servers, GPUs, and specialized control devices, has a significant carbon footprint. The disposal and recycling of hardware at the end of its life cycle also have environmental impacts. Sustainable practices in hardware lifecycle management can reduce the overall carbon footprint.
\end{itemize}

Mitigating this impact and reaching carbon neutrality has been a field of research going further than the field of AO for astronomy. First, the efficiency of the algorithms used in RTC systems can significantly impact energy consumption. More efficient algorithms require less computational power and can reduce the overall energy demand. Optimizing the code for energy efficiency, such as reducing unnecessary computations and improving data handling, can also contribute to lowering the carbon footprint\cite{gratadour2012}. Furthermore, integrating renewable energy and improving cooling technologies within observatories would play a key role. Along these lines, the use of shared-memories can play a role with respect to the data management: solving the PSF reconstruction challenge would allow for considerably reducing the amount of data being stored.

Further work is necessary to accurately assess the carbon footprint of RTC, and how to mitigate that to reach carbon neutrality.


\section{Conclusion}
This proceeding covers an introduction to the realm of RTC for users and WFS\&C developers, and present a form of step-by-step guide on how to install and interact with a RTC. 
Along with this intention, the purpose is equally to encourage the RTC community to adopt common software standard and practices for increased collaboration. 
Collaborative development across various instruments and telescopes benefits significantly from standardized protocols and shared-memory architectures, which allow for more straightforward adaptation and integration of new features. 
Indeed, gathering several contributing teams and a diversity of experiences will promote and fast-track the development of RTCs, enabling passing the tipping point for a self-sustaining community of astronomers and instrument scientists. 

Finally, successful RTC integration within an AO system's cyber infrastructure involves harmonizing with observatory systems, enhancing user experience through an effective GUI, and ensuring reliable feature integration through meticulous calibration, testing and version control. These practices collectively contribute to the efficient and effective operation of high-contrast imaging systems for exoplanet detection and characterization.



\acknowledgments 
The development of SCExAO is supported by the Japan Society for the Promotion of Science 
(Grant-in-Aid for Research numbers 23340051, 26220704, 23103002, 19H00703, 19H00695 and 21H04998), 
the Subaru Telescope, the National Astronomical Observatory of Japan, the Astrobiology Center of the National Institutes of Natural Sciences, Japan, the Mt Cuba Foundation and the Heising-Simons Foundation.
The authors wish to recognize and acknowledge the very significant cultural role and reverence that the summit of Maunakea has always had within the indigenous Hawaiian community, and are most fortunate to have the opportunity to conduct observations from this mountain.

\bibliography{report} 
\bibliographystyle{lyotspiebib} 

\end{document}